\documentclass[5p]{elsarticle}

\usepackage[latin1]{inputenc}
\usepackage{amsmath}
\usepackage{amsfonts}
\usepackage{amssymb}
\usepackage{epsfig, caption}
\usepackage{xcolor,soul}
\sethlcolor{yellow}
\definecolor{orange}{rgb}{1.0, 0.5, 0.0}

\begin{document}
\title{Anharmonic contributions in real RF linear quadrupole traps}

\author[]{J.~Pedregosa\corref{cor1}}
\ead{jofre.pedregosa@univ-provence.fr}
\cortext[cor1]{Corresponding author}
\author[]{C.~Champenois}
\author[]{M.~Houssin}
\author[]{M.~Knoop}

\address{Physique des Interactions Ioniques et Mol\'eculaires (UMR 6633), CNRS - Aix-Marseille Université, Centre de Saint J\'er\^ome, Case C21, 13397 Marseille Cedex 20, France}

\begin{abstract}
The radiofrequency quadrupole  linear ion trap is a widely used device in physics and chemistry. When used for trapping of large ion clouds, the presence of anharmonic terms in the radiofrequency potential limits the total number of stored ions. In this paper, we have studied the anharmonic content of the trapping potential for different implementations of a quadrupole trap, searching for the geometry best suited for the trapping of large ion clouds. This is done by calculating the potential of a real trap using SIMION8.0, followed by a fit, which allows us to obtain the evolution of anharmonic terms for a large part of the inner volume of the trap.
\end{abstract}

\maketitle

\section{Introduction}
The trapping of charged particles by radiofrequency (RF) electric
fields was first demonstrated in 1954 \cite{paul90} and quickly
proved to be an  extremely powerful tool for the experimental
investigation of a wide range of phenomena. In particular the linear
quadrupole RF trap is found at the heart of many experiments where
few to many ions are laser-cooled  to very low temperature for
applications in optical frequency metrology \cite{rosenband08},
quantum computation ~\cite{haeffner05,hume07} and formation of large
Coulomb crystals~\cite{drewsen98}. In this paper, the deviation from
the ideal harmonic potential is evaluated for different
implementations of linear quadrupole ion traps, in order to maximize the number of trapped ions.

\begin{figure}
\centering
\hspace{10mm} 
\includegraphics[width=0.4\textwidth]{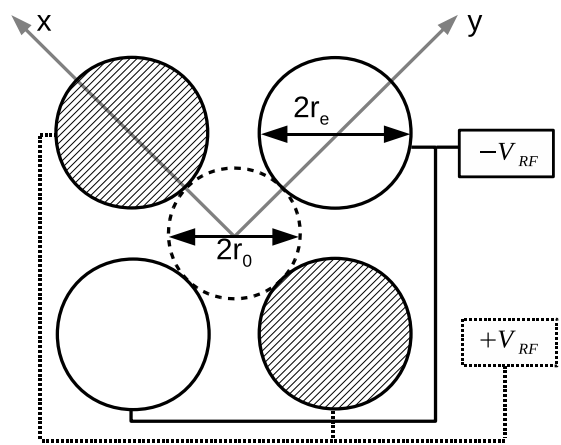}
 \caption{Schematic transverse view of a linear RF trap made of four cylindrical electrodes. The  axes convention is shown, as well as the definitions of the applied RF voltages $V_{rf}=\frac{U_0 + V_0\cos{\Omega t}}{2}$, the trap radius $r_0$ and the electrode radius $r_e$.}\label{fig:figure1}
\end{figure}

In this article, we briefly recall the basic information about
linear quadrupole traps; detailed descriptions of RF traps and  trajectories of the stored ions can be found in several textbooks
(e.g.~\cite{March95}). The most simple linear RF trap is formed by
four parallel rods of radius $r_e$, see figure~\ref{fig:figure1}. In
such devices, the trapping of charged particles in the transverse
direction is achieved by applying an alternating potential
difference between two pairs of electrodes (called the RF electrodes
in the following). The trapping along the symmetry axis of the trap
(to be taken as the $z$ axis) is realised by a static potential
applied to two electrodes sets at both ends of the trap (called the
DC electrodes in the following). The potential at the centre of a
quadrupole linear trap is then well approximated by:

\begin{eqnarray}\label{eq:ideal_case}
 \phi(x,y,z,t)  & =&    \frac{(U_0 + V_0\cos{\Omega t})}{2r^2_0\mathcal{L}} (x^2 - y^2) \nonumber \\
 &+& \frac{\kappa U_{dc}}{z_0^2}\left( \frac{ 2z^2 -x^2 - y^2}{2} \right)
\end{eqnarray}
where $V_0$ is the amplitude of  the RF potential difference between
two neighbouring electrodes, $U_0$ a static contribution to this
potential and $U_{dc}$ the voltage applied to the DC electrodes.
$\Omega$ is the angular frequency of the RF voltage, $r_0$ is the
inner radius of the trap, $2 z_0$ is the length
of the trap, corresponding to the distance between the
DC electrodes, $\kappa$ and $\mathcal{L}$ are  geometric loss factors. $\kappa$ is mainly induced by shielding effects whereas $\mathcal{L}$ stands for a loss in trapping efficiency due to the non-ideality of the quadrupole \cite{schrama93}.  While in most cases $\phi$ is an accurate approximation close to the trap
axis and around $z\simeq0$, strong deviations occur
further away from the center. The deviations induced depend strongly on the geometry
used for the trap, as it is shown in the present article.

The motion of a single ion in a potential defined  by
$\phi(x,y,t)$ is described by the Mathieu differential
equations \cite{McLachlan47}. The stability of the trajectory is
controlled by the Mathieu  parameters, which depend on the working
parameters of the trap and on the charge ($Q$) and mass ($M$) of the
particle:
\begin{align}
    & q_x = -q_y = \frac{2QV_0}{M\Omega^2 r_0^2\mathcal{L}} \nonumber \\
    & a_x = a-\Delta a ~~;  & a_y = -(a+\Delta a)  \\
    & a = \frac{4QU_0}{M\Omega^2 r_0^2\mathcal{L}} ~ ; &\Delta a = \frac{4\kappa Q U_{dc}}{M z_0^2 \Omega^2} \nonumber
    \end{align}
where $a$ represents the Mathieu parameter of the  pure quadrupole
mass filter, and $\Delta a$ is the perturbation  introduced by the
presence of the axial confining voltage~\cite{drewsen00}. Most traps
are operated in the adiabatic regime, where it is possible to
decouple the short time scale of the RF driven motion, called
micromotion, and the longer time scale of the motion induced by the
envelope of the electric field, known as
macromotion~\cite{dehmelt67}. In this frame, the ion's motion is the
superposition of a periodic oscillation in an effective harmonic
potential, characterized by the radial frequencies  $\omega_x(a_x,
q_x, \Omega)$ and $\omega_y(a_y, q_y, \Omega)$, with the RF-driven
motion (at frequency $\Omega$) with an amplitude proportional to the local RF electric
field.

The simultaneous trapping of two or more ions  couples the $x$ and
$y$ equations of motion through a nonlinear term induced by the
Coulomb interaction and an analytical solution does not exist any
longer. Moreover, due to this nonlinear contribution the ions gain
energy from the RF field. This energy gain depends on the ion
density and/or neutral background pressure inducing
collisions~\cite{ryjkov05}. While it can be negligible for very
dilute ion clouds in good vacuum conditions, it can be a critical
point for the storage of dense clouds where an important fraction of
the ions can be lost due to this gain of energy.

Nevertheless, even in the case of a single  ion in a perfect vacuum,
nonlinearities can be introduced in the equations of motion by
differences between the real RF and DC potentials and the ideal
case. Imperfections in the potentials come from misalignments of the
different electrodes and differences between the electrode surfaces
from the ideal equipotential surfaces. These effects are very well
known for hyperbolic Paul traps where holes in the electrodes are
necessary for the injection and extraction of ions, electrons or
light. Moreover, these undesired contributions to the potential can
contain higher harmonic contributions which lead to a special form
of RF-heating. More precisely, it has been shown~\cite{wang93} that
for an imperfect Paul trap, nonlinear resonances leading to the loss
of ions occur in the otherwise stable region,  if the ion's secular
frequencies and the radiofrequency are related by a linear
combination with integer coefficients ($N_x, N_z, k$):
\begin{equation}
    N_x \omega_x + N_z \omega_z = k \Omega.
\end{equation}
This resonant phenomena due to nonlinear couplings has been observed
experimentally  by several groups \cite{alheit95,vedel98}.

Ideal harmonic potentials  are obtained only if the shape of the
RF electrodes follows the theoretical equipotential surface, a
hyperbolic surface in the case of a quadrupole trap. The difficulty
in machining such surfaces can be overcome by
realizing circular rods. In this case, the relation
$r_{e}=1.14511r_{0}$ between the trap radius $r_0$ and the electrode
radius $r_e$ guarantees a minimal contribution of the lowest order
term in the perturbation potential expansion for infinitely long
RF electrodes \cite{reuben96}. This radius ratio is used for all the traps studied in this article. While some work has been done in optimizing
miniature~\cite{schrama93} and cylindrical~\cite{wu05} traps
designs, to our knowledge no study has concentrated on the
study of the anharmonic terms in a real linear trap.

Our research project aims at trapping a very large and laser-cooled
ion cloud to  study fundamental phenomena related to their dynamics
and thermodynamics. In order to reach a very high temporal stability
of the ion number, the trapping potential has to be quasi-ideal over
a large part of the trapping volume. The ion cloud is expected to
fill the trap to half the trap size in both directions ($- z_0/2 <
z< z_0/2$ and $ 0 <r< r_0/2$) with a sufficiently high ion density
such that a high number of ions ($\sim 10^7$) can be reached in a
compact trap.

The present study evaluates the anharmonic contributions to the
trapping potential of various existing devices and proposes an
optimised alternative version. Contributions due to the finite size
of the RF electrodes  and  to the shape of the DC electrodes are
quantified and compared. Contrary to former studies in spherical traps \cite{schrama93} where anharmonic contributions were
evaluated in the very centre of the trap ($r/r_0, z/z_0 < 0.1$),
which is relevant for single ions or chains of single ions, the
present study is made in three dimensions and extends to values
beyond $z_0/2$. This is motivated because a large ion cloud explores
a much bigger volume of the trap, and therefore it is desirable that
this volume has a low anharmonic content in order to minimise
RF-heating.

\begin{figure}
 \centering
\hspace{10mm}\includegraphics[ width=0.5\textwidth] {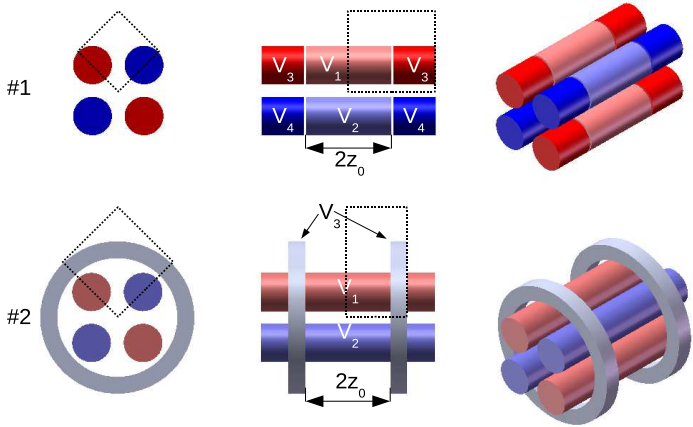}
\caption{(Color Online)~Design of the first two analysed geometries. The dimensions used in both cases are: $r_0=4$ mm and $z_0=10$ mm, with $r_e=1.14511r_0$. Trap \#1 is formed by four segmented rods whose electrical connections are detailed in table~\ref{tab:config_elec}. Trap \#2 consists of four continuous rods surrounded by DC electrodes made of two external rings of longitudinal dimension $l_{ring} = 4$ mm, inner radius $r_{int} = (r_0 + 2 r_e) + 1$ mm  and outer radius $r_{ext} = r_{int} + 5$ mm. The distance between these rings  defines $2z_0$. The dotted lines define the volume used for the SIMION8.0 calculations. }
\label{fig:figure2}
\end{figure}

Figure~\ref{fig:figure2} shows the two  geometries initially
analysed. We have chosen only devices which leave the z-axis free as
this axis is often needed to implement laser cooling  and/or to
introduce/extract ions. The segmented rod geometry (\#1) is used,
among others, by the Ion Trap Group in {\AA}arhus University to trap
large ion crystals~\cite{drewsen98,drewsen00}.

This geometry can be used with  two different electrical
configurations, named \#1.1 and \#1.2 (see
table~\ref{tab:config_elec} for details).  \#1.2 has been analyzed
due to its simplicity as \#1.1 requires a much more advanced
electronic configuration than \#1.2 (see
table~\ref{tab:config_elec}). The second implementation, \#2, is
similar to  one of the trap used in Quantum Optics and Spectroscopy Group at
Innsbruck University for quantum information
experiments~\cite{naegerl98}.

\begin{table}\begin{center}
 \caption{Electrical configuration for the different trap geometries. }
    \begin{tabular}{|c | c | c | c | c |}
        \hline
               & $V_1$             & $V_2$             & $V_3$  &                $V_4$\\
        \hline
        \#1.1: &  $ +V_{rf}$  &  $ -V_{rf}$ & $ U_{dc} + V_{rf}$ & $ U_{dc} - V_{rf}$ \\
        \#1.2: &  $ +V_{rf}$ &  $ -V_{rf}$ & $ U_{dc}$ & $ U_{dc}$ \\
        \hline
        \#2:   &  $ +V_{rf}$ &  $ -V_{rf}$ & $ U_{dc}$ & -- \\
        \hline
    \end{tabular}

    \label{tab:config_elec}
\end{center}\end{table}

This article is organised  in several sections. The method used to
calculate the potential created by a given geometry and the fitting
procedure is described in section \ref{sec:comppot}. In section
\ref{sec:anh_contr}, the anharmonic contributions are calculated and
evaluated for the RF-part while we discuss the DC potential in
section~\ref{sec:dc_pot}. Finally, an alternative geometry is
introduced and analysed in section \ref{sec:hybr_geom}, followed by
the conclusion, section \ref{sec:conclusion}.

\section{Fitting the ``real'' trap potentials}\label{sec:comppot}
\subsection{Computation of the potentials}
The commercial software SIMION8.0 \cite{simion} has been used to
numerically solve the Laplace equation for each geometry. SIMION8.0
uses a Finite Difference Method (FDM), where the solution is
obtained at each node of the mesh used to describe the electrodes
and the space between them. The error on the electrostatic potential
using FDM scales with $1/h$ \cite{strang73}, where $h$ is the node
density (number of nodes/mm), but the reached accuracy depends on
the particular geometry of the electrodes. In order to check the
$1/h$ dependency and to obtain an estimation of the accuracy of the
results given by the software, we have used SIMION8.0 to calculate
the potential for infinitely long electrodes~\footnote{In SIMION8.0,
it is possible to define a 2D plane only, in which case, the program
assumes this 2D section extends to $[-\infty, +\infty]$} with
hyperbolic sections in which case the potential can be exactly
described by: $\phi_{a}(x,y) = \frac{x^2 - y^2}{2r_0^2}$ when
$V_0=0$ and $U_0=1V$. Note that while the confinement is achieved
dynamically,  the anharmonic contributions to the potential do not depend on time and can be analysed as a static property of the trap.

We can  then calculate the average relative error $Er(h)$, as a
function of the node density $h$, using:
\begin{equation}\label{eq:residu_n}
    Er(h) = \frac{1}{n_x n_y}\sum_{i}^{n_x} \sum_{j\ne i}^{n_y}{ \left|\frac{\phi_{s}(x_i,y_j) - \phi_a(x_i,y_j) }{\phi_a(x_i,y_j)} \right|},
\end{equation}
where  $\phi_{s}(x_i,y_j)$ is the SIMION8.0 solution with $x_{i+1} -
x_{i} = y_{i+1} - y_{i}= h$ and $n_x$, $n_y$ are the number of
nodes.

\begin{figure}
 \centering
\hspace{10mm}\includegraphics[ width=0.5\textwidth] {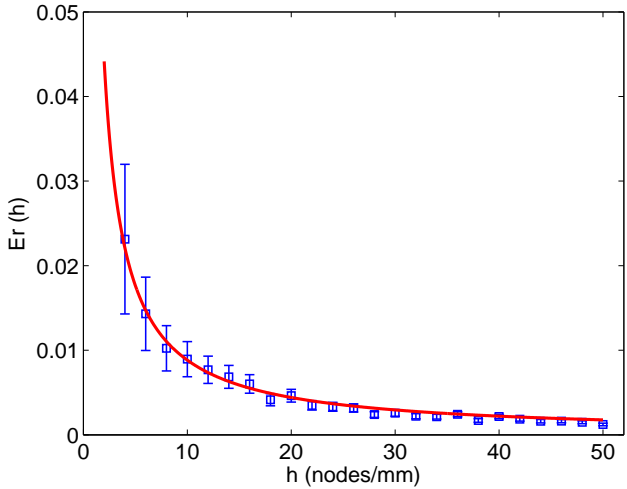}
 \caption{(Color Online)~Estimation of the error between the numerical solution given by SIMION8.0 and the known analytical potential in the case of infinite hyperbolic electrodes as a function of the mesh density $h$. The dots show $E_r(h)$ calculated by equation~\ref{eq:residu_n} whereas the curve plots $f(h)=a /h$, with $a = 8.8 \times10^{-2}$. The error bars correspond to $1\sigma$. }
\label{fig:figure3}
\end{figure}

Figure~\ref{fig:figure3} shows $Er(h)$ for $0 \leq x \leq r_0/2$ and
$0 \leq y \leq r_0/2$ (the axes are oriented as in
Fig~\ref{fig:figure1}). %
Fig~\ref{fig:figure3} also shows a fit using
$f(h)=a/h$. The comparison of the two plots indicates that $Er(h)$
closely follows the $1/h$ law. The error bars correspond to one
standard deviation, $\sigma$.

For the computation of  the potential created by \#1 and \#2, all
possible symmetries have been used to reduce the required volume, as
indicated in figure~\ref{fig:figure2} by the dashed lines. This is
important due to computer memory issues, which  limit the mesh
density which can be used for a given problem. In the present study,
we were limited to a node density of 32 nodes/mm, for all computations carried out for a 3D volume.
Figure~\ref{fig:figure3} shows that this mesh density already
provides an accuracy of 0.3 \% for the hyperbolic trap. As the sizes of the studied traps are identical to the hyperbolic trap, the  precision reached by SIMION8.0 is of the same order.
Therefore, the use of the potentials computed by SIMION8.0 in the study of the
anharmonic content in a quadrupole linear trap is fully justified.

\subsection{Fitting the potentials}
An analytical two-dimensional solution of the Laplace equation for a
quadrupole trap with circular sections exists, if we assume
infinitely long RF electrodes~\cite{reuben96}:
\begin{equation} \label{eq:anal_phi}
    \phi(x,y) = C_0+Real\left\{ \sum^{\infty}_{m=0}{C_{(4m + 2)} \xi^{(4m + 2)} }\right\}; ~~~~~~ \xi = x + iy
\end{equation}
where $C_{(4m+2)}$ are real coefficients  and we have introduced a
uniform contribution $C_0$ to  equation (\ref{eq:zeta}) of~\cite{reuben96}.

\begin{figure}
 \centering
\hspace{10mm}\includegraphics[ width=0.52\textwidth] {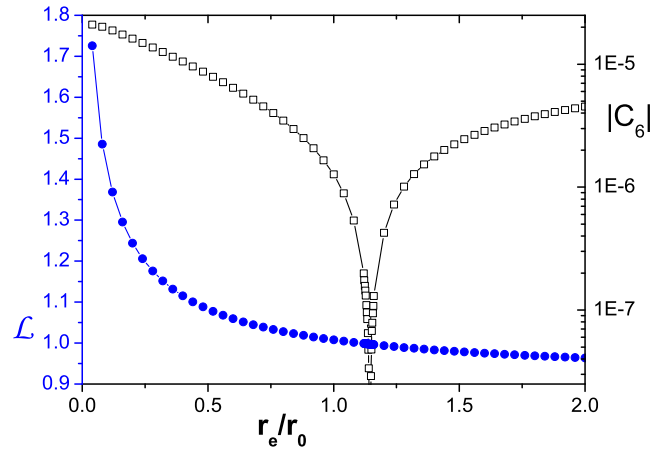}
\caption{(Color Online)~Squares and right axis:  absolute value of the first anharmonic coefficient $C_6$, versus $r_e/r_0$ for $r_0=4$~mm.  Blue circles and left axis: loss factor, $\mathcal{L}$ versus $r_e/r_0$. } \label{fig:figure_ext}
\end{figure}

As pointed out in the introduction, the term $C_6$ vanishes for infinitely long electrodes for  $r_e=1.14511r_0$ \cite{reuben96}. To test our analysis of the potentials, we have computed the potential created by a 2D mass filter (infinitely long RF electrodes, without DC electrodes) for several values of the radius ratio $r_e/r_0$ and found the $C_{(4m+2)}$ coefficients of equation~\ref{eq:anal_phi}, by a fitting procedure analysed in the following. The evolution of $|C_6|$ as a function of $r_e/r_0$ shown in figure~\ref{fig:figure_ext} shows  that the $C_6$ term can be decreased by 3 orders of magnitude by choosing the appropriate radius ratio and confirms the zero crossing of $C_6$  for a $r_e/r_0$ value between $1.144$ and $1.146$. This accuracy corresponds to the maximal spatial resolution of our mesh for this problem (a high mesh density of 500~nodes/mm was possible due to the 2D nature of the problem). While this method does not provide a precise estimation of the ``magic'' radius ratio, it clearly agrees with the analytical result of $r_e=1.14511r_0$, showing that SIMION8.0 produces a very good representation of the potential created by an RF-trap. Higher order terms show a similar quantitative behavior as $C_6$, with respective minima for $C_{10}$ at $r_e/r_0 \approx 0.78$, and $r_e/r_0 \approx 0.58$  for $C_{14}$.

In addition, figure~\ref{fig:figure_ext} shows the ratio between the ideal and the computed $C_2$ coefficients which corresponds to the geometrical loss factor, $\mathcal{L}$ introduced in equation~\ref{eq:ideal_case}:
\begin{eqnarray}
	\mathcal{L} & = & \frac{C_2^{ideal}}{ C_2^{real} }  \\
	C_2^{ideal} & = & \frac{U_0}{2r^2_0} \nonumber
\end{eqnarray}
For the chosen radius ratio, there is no loss of trapping efficiency induced by the circular shape of the electrodes ($\mathcal{L} \simeq 1$).  We can observe that $\mathcal{L}$  takes values below 1 for larger electrode radius,  which for experimental reasons (decreasing inter-electrode distance) might not be an appropriate choice. In the following, all the studied trap configurations have the same radius $r_0=4$~mm and $r_e=4.58$~mm.

The finite size of the rods and the presence of the DC electrodes
introduce a dependency on $z$ of the coefficients $C_{(4m+2)}$. This
dependency can be obtained by performing a fit at each node along
$z$ to the numerical solution obtained by SIMION8.0. The trap
parameters chosen are $V_0=0$, $U_0=1V$ and $U_{dc}=0$. These
values remain the same throughout the paper, unless specified
otherwise. A truncated version of
eq.\ref{eq:anal_phi} is used to perform a least-squares
fit~\cite{bevington69}. The points included in the fitting procedure belong to  a 2D-square defined by $0 \leq x \leq r_0/2$ and $0 \leq y \leq r_0/2$, as $r = r_0/2$ is
considered as the maximum expected radius of the trapped ion cloud.

\begin{figure}
 \centering
\hspace{10mm}\includegraphics[ width=0.5\textwidth] {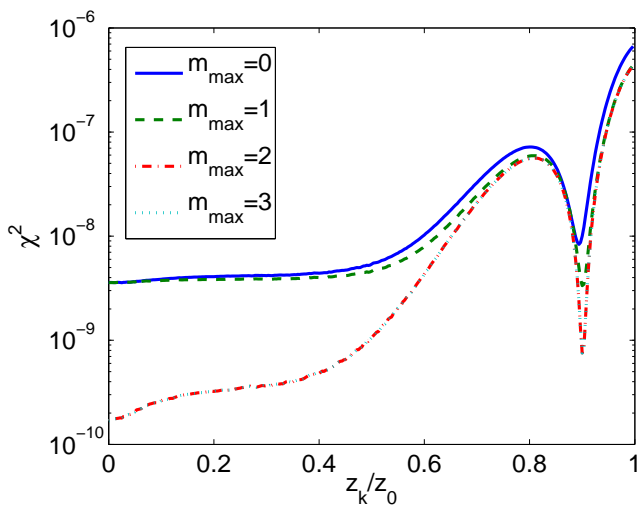}
\caption{(Color Online)~Evolution of the $\chi^2$ of the fitting procedure versus $z_k/z_0$, for increasing highest order $m_{max}$ included in the fitting equation \ref{eq:anal_phi}. The curves corresponding to $m_{max}$ = 2 and 3 overlap each other, showing that no gain is obtained by retaining orders higher than $m_{max}=2$.} \label{fig:figure4b}
\end{figure}

To demonstrate the relevance of the fit, the $\chi^2$ of the fitting procedure for geometry \#1.1 is plotted on figure \ref{fig:figure4b} for each $z_k$ and for increasing maximum order $m_{max}$ included in the fitting equation (\ref{eq:anal_phi}).
These curves illustrate several features. First, they confirm that the $C_6$ ($m=1$) contribution does   not modify the quality of the fit in the center of the trap, which is expected from the choice of the electrode radius $r_e$ that nulls the $C_6$ contribution to the potential for infinitely long electrodes. Second, they demonstrate  that adding higher orders than $m=2$ does not lead to an improvement of the fit  and in the following $m=2$ is the highest order retained in the fitting function. Furthermore, the $\chi^2$ parameter, which is lower than $10^{-9}$ for $z_k \leq z_0/2$  and $m_{max}=2$ increases drastically from $z_k = z_0/2$ to $z_0$. From $z_k/z_0 = 0.6$, one can see that adding higher order terms to the fitting equation does not result in an improvement of the $\chi^2$ parameter. This means that equation (\ref{eq:anal_phi}) is not relevant in this part of the trap to reproduce the potential calculated by SIMION8.0 and that another analytical expression should be introduced to represent the potential close to the DC electrodes. As we are interested in  fitting the potential for $z_k/z_0 < 0.5$, curves of figure \ref{fig:figure4b} confirm the relevance of the fitting equation on the volume of interest.

To compare the precision of the fitting procedure to the precision of the SIMION8.0 calculation, we define the relative
 error of the fit, $Er_{fit}(z_k)$, taken as the  relative difference averaged over the transverse plane, between the SIMION8.0 solution,
$\phi_{s}(x_i,y_j,z_k)$, and the fitted potential, $\phi_f(x_i,y_j,z_k)$:
\begin{equation}\label{eq:residu}
    Er_{fit}(z_k) = \frac{1}{n_x n_y}\sum_{i}^{n_x} \sum_{j\ne i}^{n_y}{ \left| \frac{\phi_{s}(x_i,y_j,z_k) - \phi_f(x_i,y_j,z_k)}{\phi_{s}(x_i,y_j,z_k)}\right| }
\end{equation}
 The relative error $Er_{fit}(z_k)$ is calculated for the geometry \#1.1 and \#2 for $z_k \leq z_0/2$.  For this relative error, the addition of the $m=2$ contribution to the fitting potential allows reducing  $Er_{fit}(z_k)$ by a factor of 3 in  the centre of the trap and by a factor of 2 at $z_k=z_0/2$. The relative error then reaches $5\times 10^{-6}$ in the center of the trap for both geometries and $8\times 10^{-6}$ for geometry \#2 and $13\times 10^{-6}$ for geometry \#1.1 at $z_k=z_0/2$. This relative error is three orders of magnitude smaller than the relative error one can expect from the SIMION8.0 calculations. Nevertheless, fitting the calculated potential  is relevant for comparison between different geometries as the relative precision of the fit is as good for the two geometries considered in the text and the precision of the SIMION8.0 calculations should be identical  as the basic geometry is the same for all the configurations compared here. Once all the coefficients $C_{4m+2}(z_k)$ are determined at each
$z_k$ and for each configuration, it is possible to study the evolution
of the anharmonic contribution along the trap axis and to compare the different configurations.

\section{Anharmonic contribution}\label{sec:anh_contr}
The relative anharmonic contribution, $\zeta(x,y,z)$ of the potential in the trap can be obtained using:
\begin{equation}\label{eq:zeta}
     \phi_r = \phi_{q}(1+\zeta +\Delta\zeta)
\end{equation}
with
\begin{eqnarray*}
 \phi_{q} &=& C_0 + Real\{C_2\xi^2\} \\
     \zeta &=& \frac{Real\{C_6\xi^6 + C_{10}\xi^{10}\}+\Delta\phi_f}{C_0 + Real\{C_2\xi^2\}}  \\
    \Delta\zeta &=&  \frac{ \Delta\phi_s}{C_0 + Real\{C_2\xi^2\}}
\end{eqnarray*}
where $\phi_r$ represents the real potential created by each geometry and $\phi_q$ its harmonic contribution. $ \Delta\phi_f$ is the difference between the SIMION8.0 solution and the fitted potential calculated using the  $C_{4m+2}(z_k)$ coefficients found by the fitting algorithm.  $\Delta\phi_s$ represents the unknown difference between the real and the SIMION8.0 solution, $\Delta\zeta$ is its relative value with respect to the harmonic contribution $\phi_{q}$.

In order to facilitate the visualisation of the behaviour of the
anharmonic content, it is useful to calculate the averaged absolute value of $\zeta$
along the $z$-axis:
\begin{equation}\label{eq:zeta_avg}
    \langle\zeta(z_k)\rangle = \frac{1}{n_x n_y}\sum_{i}^{n_x} \sum_{j}^{n_y}{ \left| \zeta(x_i,y_j,z_k) \right| }
\end{equation}

\begin{figure}
 \centering
\hspace{10mm}\includegraphics[ width=0.5\textwidth] {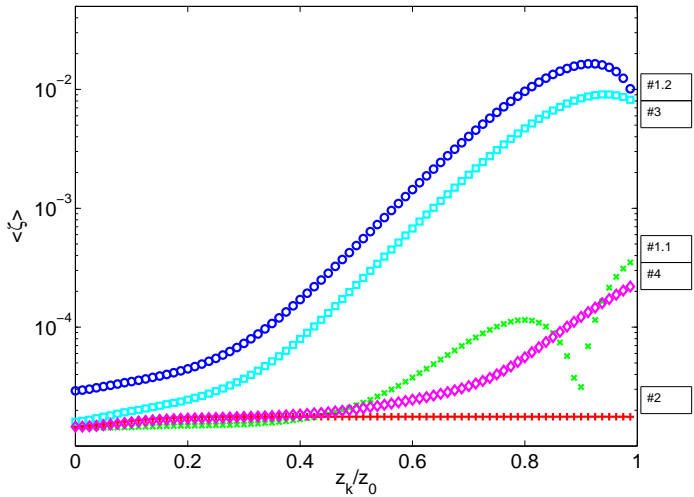}
\caption{(Color Online)~Averaged anharmonic contribution $\langle\zeta(z_k)\rangle$ along the trap axis for the configurations \#1.1, \#1.2, \#2. Configurations \#3 and \#4 are introduced later in the text. Strong differences between the trap configurations appear from $z_k/z_0>0.2$. For clarity, only every $4^{th}$ node is plotted. }
\label{fig:figure5}
\end{figure}
The evolution of $\langle\zeta(z_k)\rangle$ is shown in
figure~\ref{fig:figure5} for the configurations \#1.1, \#1.2 and \#2. While all the configurations have comparable
anharmonic content at the centre of the trap,  strong
differences appear between them when leaving the trap centre. As expected, due to the length of the RF electrodes with respect to the studied volume, configurations \#1.1
and \#2 show lower values
of $\langle\zeta(z_k)\rangle$ compared to \#1.2.  The increment observed from $z_k/z_0 > 0.5$ in the
case of \#1.1  can be explained  by   the unavoidable gap between RF and DC
electrodes which generates a rise of the anharmonic content
at the edge of the trap. The dip observed for
$\langle\zeta_{\#1.1}(z_k)\rangle$ is due to  a zero crossing of the $\zeta(x_i,y_j,z_k)$ in equation(\ref{eq:zeta_avg}). In the area of interest $\langle\zeta(z_k)\rangle$ remains rather constant along the trap axis for \#2 and \#1.1.

Figure~\ref{fig:figure5}  shows that the truncation of the RF electrodes in configuration \#1.2 leads to a substantial increase of the anharmonicity from $z_k/z_0 > 0.3$, which could limit the usable trapping
volume. To demonstrate the effect of the boundary conditions, we have calculated the potential created by a
configuration called \#3, similar to \#1.2 but with the DC-rods removed. We observe in fig ~\ref{fig:figure5} that
this leads to a reduction of the the anharmonic terms,
illustrating the importance of geometric aspects.

We  emphasize here  that the only difference between \#1.1 and \#1.2 is in the
voltage applied to the DC electrodes. Therefore, the discretization
of the mesh by SIMION8.0 and so the calculation error is exactly the same for the two configurations. The fact that the
expected behaviour of the anharmonic contribution  is reproduced in each case
re-enforces the validity of the assumption that SIMION8.0 error contribution is the same for all the configurations and that the comparative study performed in this
article is significant.

\section{DC-potentials}
\label{sec:dc_pot} 
\begin{figure}
\centering
\hspace{10mm}\includegraphics[ width=0.5\textwidth] {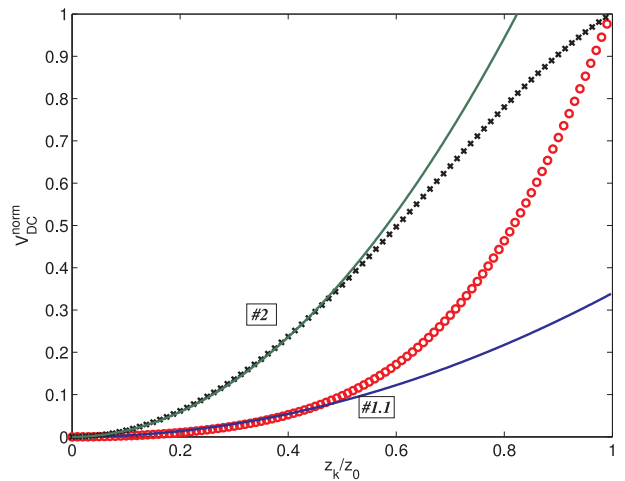}
\caption{(Color Online)~Normalised potentials created by the DC electrodes on axis (0,0,z) for geometries \#1 (empty circles)  and \#2 (crosses) together with their respective fits to $f(z)=az^2$(solid lines). Notice that the fit was performed considering only $V( z_k \le 0.5 z_0)$.}
\label{fig:figure7}
\end{figure}

In this  section, we focus on the potential
created by the DC electrodes inside the trap, as it also strongly
depends  on the chosen configuration. While at  the very centre of the trap, the potential  can  be
approximated by an harmonic function, it is not necessarily true if
larger volumes of the trap are considered. A harmonic potential shape can be
desirable as it is often a necessary assumption when describing the
ion dynamics  or equilibrium properties inside the trap, as for example to calculate the aspect
ratio (radius over length) of an ion cloud~\cite{turner87}.
Figure~\ref{fig:figure7}
shows the
potentials created only by the DC electrodes ($V_0=U_0 = 0$) of the
geometries \#1 and \#2, along the z-axis. For comparison, these potentials have been normalised using:
\begin{equation}\label{eq:normalization}
    V^{norm}_{DC}(x,y,z) = \frac{V_{DC}(x,y,z) - V_{DC}(0,0,0) }{V_{DC}(0,0,z_0)-V_{DC}(0,0,0)}.
\end{equation}
A fit to $f(z)=az^2$ using only the values of the potential in the interval $0 \leq z_k/z_0 \leq 0.5$ (see fig~\ref{fig:figure7}) clearly shows the divergence from an harmonic potential at the outer edges of the trap. In order to quantify the degree of harmonicity, we successively extended the fit, starting from the centre, until the correlation coefficient $R^2$, commonly used to quantify  the goodness of a fit, was worse than $0.9990$. This was obtained at $z_k/z_0 = 0.240$ and $z_k/z_0 = 0.597$ for \#1 and \#2 respectively.

Another issue to take into consideration is the screening of the DC-potential due to the presence of the RF electrodes, which is masked in figure~\ref{fig:figure7} due to the normalisation used. Assuming a potential depth $\Delta V = V(0,0,z_0)- V(0,0,0)$ of 1 V, we would need to apply to the DC electrodes 3.1 V for geometry \#1, but 5.5 kV for \#2. The requirement for very high voltages, without being critical in many situations, can make option \#2 inappropriate for an experimental setup where a high potential depth is needed or when designing an ion trap for space applications where the weight of high power supplies eliminates this configuration. On the other hand, \#1.1 presents the disadvantage of the complexity of the electronics needed for driving the electrodes of the trap.

The methodology presented so far, allows to  study alternative designs and evaluate their performance. As a consequence, we propose an alternative geometry, which is introduced and analysed in the next section, showing that it is possible to achieve similar performances concerning low anharmonic contributions as \#1.1 and \#2, without the need for high voltages or advanced electronics.

\section{An alternative geometry}
\label{sec:hybr_geom}
\begin{figure}
 \centering
\hspace{10mm}\includegraphics[ width=0.5\textwidth] {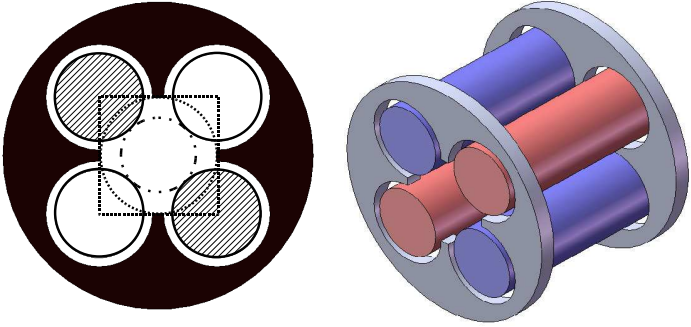}
\caption{(Color Online)~Description of the proposed alternative geometry, \#4. The electric connections are the same as for trap \#2, see table~\ref{tab:config_elec} for details. The DC electrodes geometry has been chosen so that the inner cut corresponds to the circumference inscribed in a square defined by the four rod centres.}
\label{fig:figure8}
\end{figure}

An alternative geometry, denoted as \#4 in the following, is described in fig~\ref{fig:figure8}. It uses continuous rods as in configuration \#2 but with a different design of the DC electrodes that strongly decreases the screening of the RF electrodes.
The evolution along the $z$-axis of the averaged anharmonic contribution $\langle\zeta(z_k)\rangle$  is shown in Fig~\ref{fig:figure5}. We observe that the anharmonic content is comparable to the configuration \#2 up to $z_k/z_0 \sim 0.4$ and lower than \#1.1 for the rest of the trap.

Regarding the DC-voltages, the correlation coefficient $R^2$ is found to be  smaller than $0.9990$ for $z_k/z_0 \leq 0.28$. Although the degree of harmonicity on the axial potential is not as high as for trap \#2, this alternative geometry only needs an applied voltage of 37~V in order to have $\Delta V = 1$~V. For some experiments, a smaller voltage can be preferable than a perfect harmonic behaviour on the axial direction.

It is worth mentioning that the inner cut off of the DC electrodes corresponds to the circumference inscribed in a square centred on the electrodes as shown in figure~\ref{fig:figure8}. A geometry with the inner radius cut off equal to $r_0$ was also studied. The voltage required in that case for a potential depth $\Delta V = 1$~V was further reduced to 3.8~V. However, this reduced shielding is paid by an increase of the anharmonicity which reaches the level of the worse geometry \#1.2. This comparison shows that the anharmonic contributions are very sensitive to the position of the inner cut-off radius  and configuration \#4 is a good solution to reduce these contributions.

\section{Conclusion}\label{sec:conclusion}

The anharmonic contribution to the potential created by a linear quadrupole trap has been studied for a large trap volume for several implementations of the DC and RF electrodes.  We found that while the anharmonic content is similar at the centre of the trap in all the different implementations, they behave extremely differently further away. The already operated configurations \#1.1 and \#2 as well as the proposed configuration \#4 achieve similar good performances for $z_k/z_0 <0.5$.   Furthermore,  for  $z_k/z_0 >0.5$ the RF potential of configuration \#4 keeps a level of anharmonicity intermediate between  \#2 and  \#1.1. The rather complicated electronic set-up needed for the implementation of \#1.1 and the need of high static voltages for the geometry \#2, make these designs not suited for all experimental situations. The alternative geometry  \#4, proposed in this paper, represents therefore a trade-off between complexity of the implementation and low anharmoncities, and should allow the envisaged trapping of a very large ion cloud.

\section{Acknowledgements}
The authors acknowledge Fernande Vedel and Laurent Hilico for valuable comments and discussions.
This work is partly funded by CNES under contract n$^{\circ}$ 81915/00 and by ANR under contract ANR-08-JCJC-0053-01.

\end{document}